\documentclass[runningheads]{llncs}
\usepackage{graphicx}
\usepackage[bookmarks=true,bookmarksopen=true]{hyperref}

\usepackage[version=0.96]{pgf}
\usepackage{tikz}
\usepackage{pgfplots}
\usepackage{pgfplotstable}
\usepackage{listings}
\usepackage{comment}
\usepackage{wrapfig}
\usetikzlibrary{arrows,shapes,snakes,automata,backgrounds,petri}

\begin{document}

\title{Streaming Computations with Region-Based State on
       SIMD Architectures}

\author{Stephen Timcheck \and Jeremy Buhler}

\authorrunning{S.\ Timcheck \and J.\ Buhler}

\institute{Washington University in St.\ Louis \\ 
One Brookings Dr., St.\ Louis, MO 63130, USA \\
\email{\{stimcheck,jbuhler\}@wustl.edu}}

\maketitle

\begin{abstract}
Streaming computations on massive data sets are an attractive
candidate for parallelization, particularly when they exhibit
independence (and hence data parallelism) between items in the stream.
However, some streaming computations are stateful, which disrupts
independence and can limit parallelism.  In this work, we consider how
to extract data parallelism from streaming computations with a common,
limited form of statefulness.  The stream is assumed to be divided
into variably-sized \emph{regions}, and items in the same region are
processed in a common context of state.  In general, the computation
to be performed on a stream is also \emph{irregular}, with each item
potentially undergoing different, data-dependent processing.

This work describes mechanisms to implement such computations
efficiently on a SIMD-parallel architecture such as a GPU.  We first
develop a low-level protocol by which a data stream can be augmented
with control signals that are delivered to each stage of a computation
at precise points in the stream. We then describe an abstraction,
\emph{enumeration and aggregation}, by which an application developer
can specify the behavior of a streaming application with region-based
state.  Finally, we study an implementation of our ideas as part of
the MERCATOR system~\cite{MERCATOR} for irregular streaming
computations on GPUs, investigating how the frequency of region
boundaries in a stream impacts SIMD occupancy and hence application
performance.\footnote{Presented at \emph{13th Int'l Wkshp. on Programmability and Architectures for Heterogeneous Multicores, Bologna, Italy, Jan 2020}.  Copyright \textcopyright{} 2020 by Stephen Timcheck and Jeremy Buhler. All rights reserved.}

\keywords{signal \and control message \and SIMD \and irregular \and streaming}
\end{abstract}

\section{Introduction}
Streaming computations are integral to high-impact applications such
as biological sequence analysis~\cite{BLAST},
astrophysics~\cite{ASTRO}, and decision cascades in machine
learning~\cite{VIOLA}.  These computations operate on a long stream of
data items, each of which must be processed through a pipeline of
computational stages.  When items can be processed independently and
identically, extracting data parallelism is straightforward,
particularly on SIMD-parallel architectures such as GPUs.

A greater challenge lies in supporting streaming computations whose
behavior deviates from the above ideal.  Deviations can occur in two
ways.  First, items may not be processed identically; in particular,
some stages of the pipeline may produce a variable, data-dependent
amount of output for each input item.  Such computations --- which
include all the examples cited above as well as applications in,
e.g., particle simulation~\cite{barnes1986hierarchical} and network
packet processing~\cite{roesch1999snort} --- are said to exhibit
\emph{irregular} data flow, which complicates the lock-step execution
model of a SIMD-parallel processor.  Our prior work on the MERCATOR
system~\cite{MERCATOR} describes a way to support such irregular
computations on GPUs.

A second deviation, which is the subject of the present work, arises
when items in a stream cannot be processed independently; that is, the
computation is stateful.  To avoid the need for full serialization, we
focus on a common scenario in which the input stream is divided into
variably-sized \emph{regions}. Items in one region are processed
independently of each other but in a common context.  For example, a
stream of characters may be grouped into lines or network packets; a
stream of edges in a graph may be grouped by their source vertex; or a
stream of measurements may be grouped by a common time window or event
trigger.  Region boundaries are state-change events for the stream ---
items after a boundary must be processed differently than items before
it.

In this work, we investigate mechanisms to support streaming
computations with region-based contextual state on SIMD-parallel
platforms.  Our contributions are threefold.  First, we describe a
low-level mechanism for precise delivery of control signals between
pipeline stages of a streaming application.  This mechanism, unlike
those described in some prior work (e.g.\ \cite{Teleport}), supports
irregular dataflow.  Second, we use this mechanism to construct an
abstraction, \emph{enumeration and aggregation}, that lets application
developers express region-based contextual state as part of a
streaming application.  Finally, we implement our designs in MERCATOR
to investigate the performance implications of regional context,
exposing a SIMD-specific performance tradeoff between alternate ways
of implementing this behavior in applications. The two strategies we
examine trade off between SIMD occupancy and representation overhead.

The remainder of this paper is organized as follows.  Section 2
describes the application and architectural models in which we
formulate our work and considers related work in other streaming
models.  Section 3 describes our protocol for synchronizing control
signals with a data stream.  Section 4 describes the developer-facing
abstraction of enumeration and aggregation, which we implement in
terms of signals.  Section 5 investigates the performance of our
designs, while Section 6 concludes and considers future work.

\section{Background and Related Work}

\subsection{Application Model}

\begin{wrapfigure}[12]{L}{0.4 \columnwidth}
	
	\centering
	\includegraphics[width=0.25 \columnwidth]{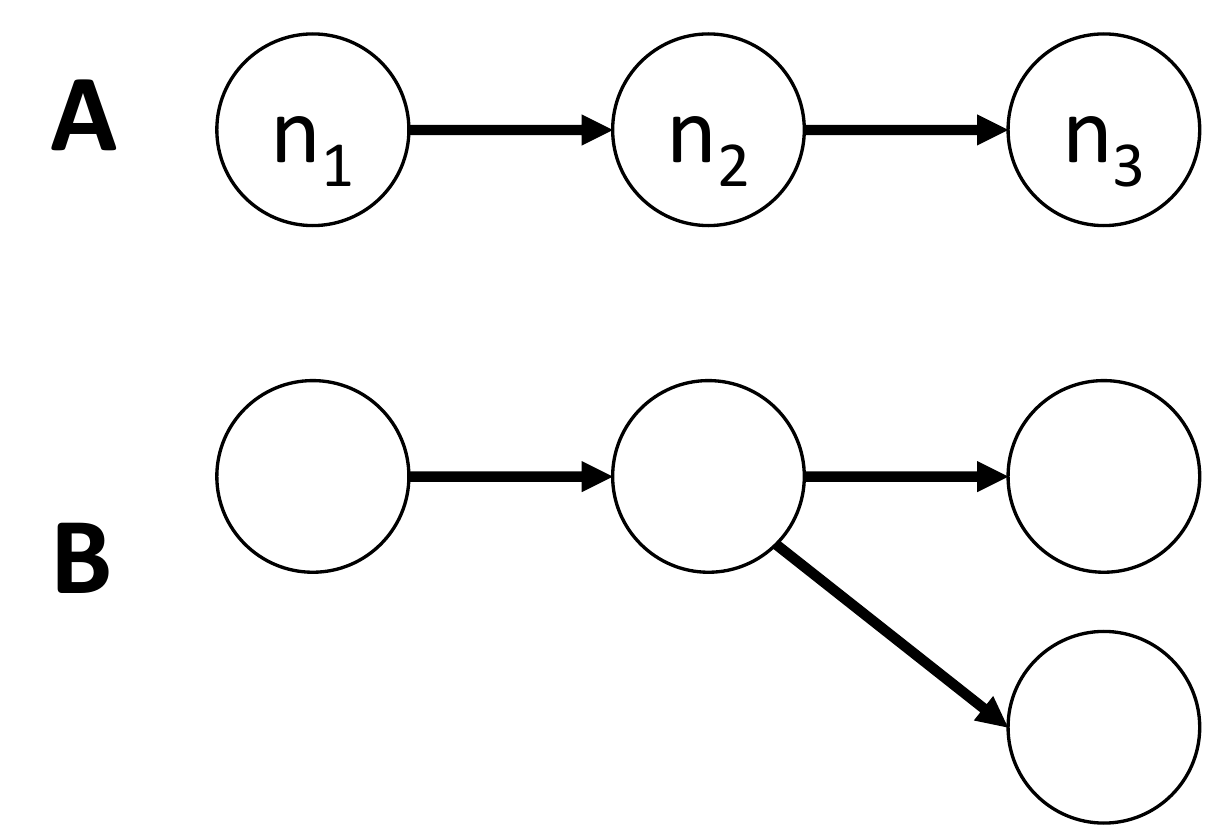}
	
	\caption{(a) Streaming computation pipeline with three compute nodes;
		(b) Pipeline with a tree topology.
		\label{fig1}}
	
\end{wrapfigure}

A streaming application consists of a pipeline of \emph{compute nodes}
connected by fixed-sized \emph{data queues}. A node consumes a stream
of \textit{data items} from its input queue and produces a stream of
data items (perhaps of a different type) at its output that are queued
for processing by the next node downstream in the pipeline. When a
node is executed to consume one or more inputs, we say that the node
\emph{fires}.  Each input data item consumed by a node causes it to
produce zero or more outputs. The number of outputs may vary for each
input consumed, up to some node-dependent maximum, and is not known
prior to execution. Figure~\ref{fig1}a shows a simple application
pipeline with three nodes.  While we mainly discuss linear pipelines
in this work, our contributions also apply to tree-structured
topologies like those in Figure~\ref{fig1}b. 

We do not consider DAG-structured topologies because the semantics
associated with convergent edges are complex under irregular dataflow,
even in the absence of stateful execution~\cite{PENG}.  Topologies
with cycles have clearer streaming semantics, but in the presence of
irregularity, items in the stream can be reordered if they take
different numbers of trips around a cycle.  For such topologies,
Maintaining precisely ordered control boundaries in the stream
requires aggressive reordering that is beyond the scope of this work.

An application is provided with an initial stream of inputs to its
\emph{source node}. When the application executes, a global
\emph{scheduler} repeatedly chooses a node with one or more pending
inputs to fire. Because of our architectural mapping below, we assume
that only one node fires at a time and that a node cannot be preempted
while firing. The scheduler continues to select and fire nodes until
no node has any inputs remaining.

A \emph{signal} is a control message generated by a node for
consumption by its downstream neighbor.  When a node receives a signal
it can change its state and may also generate additional signals to
the next node downstream.  Signals must be delivered \emph{precisely}
with respect to the stream of data items.  Formally, suppose we have
two successive nodes $n_1$ and $n_2$ in a pipeline.  If $n_1$ emits a
data item $d$, followed by a signal $s$, followed by another data item
$d'$, then $n_2$ must receive $s$ after processing $d$ but before
processing $d'$.

\subsection{Target Architecture and Mapping}

Our work targets wide-SIMD multiprocessors.  While many
general-purpose processors support SIMD instructions, we realize our
designs on NVIDIA GPUs due to their robust CUDA tool chain and their
popularity as accelerators.

A GPU may be viewed as a collection of SIMD processors sharing a
common memory. Each processor has a fixed \emph{SIMD width} $w$, which
is the number of concurrent SIMD lanes that it can execute\footnote{We
  treat the CUDA block size as the effective SIMD width, ignoring
  CUDA's virtualization of an underlying, smaller width (the warp
  size).}.  SIMD lanes execute computations in lock-step; hence,
divergent behavior such as nonuniform branches, or lack of inputs to
some SIMD lanes, causes some lanes to sit idle while others execute.

Mapping a streaming computational pipeline onto a GPU entails moving
the input stream from the host system to the GPU's memory, executing a
GPU kernel to process the data through the pipeline, and finally
transferring the output stream back to the host.  Our work focuses on
efficient processing on the GPU, rather than the orthogonal task of
host/GPU data transfer.

GPU runtimes offer little support for interprocessor synchronization
other than via control transfer back to the host.  To avoid excessive
host-device control overhead, we therefore instantiate the pipeline
separately on each processor of the GPU, then let all processors'
pipelines compete to consume data from a common input stream. This
approach requires atomic operations but no locking.  The GPU kernel
does not return control to the host until the entire stream is
consumed.  Each processor on the GPU independently implements the
sequential, non-preemptively scheduled computation described above.
However, a node now consumes not a single input per firing but rather
a variably-sized \emph{ensemble} of inputs, up to the processor's SIMD
width, which are processed in parallel.

Our performance goal is to maximize throughput, or equivalently to
minimize time for the GPU to process an entire input stream. A
secondary goal that promotes high throughput is to maximize \emph{SIMD
  occupancy}, the fraction of SIMD lanes doing useful work at any step
of the computation.  In particular, we want the sizes of ensembles
presented to each node to achieve the full SIMD width.  However, we
will show that region-based state can be in tension with the goal of
maximizing SIMD occupancy, creating a challenge for performance
optimization.

\subsection{Related work}

Several systems and languages have been developed to express streaming
dataflow computations.  One of the most influential such systems is
StreamIt~\cite{STREAMIT}, which implements the synchronous data flow
(SDF) abstraction~\cite{SDF}.  StreamIt assumes a fixed number of
outputs per input to a compute node.  This assumption allows StreamIt
to offer a powerful abstraction, \emph{teleport
  messaging}~\cite{Teleport}, in which signals can flow both forward
and backward in a pipeline.  StreamIt can also schedule a signal to be
delivered to an arbitrary destination node at some precise future
time, rather than forcing the signal to flow through the pipeline.

Many capabilities of teleport messaging rely on the underlying SDF
model.  In contrast, our model does not assume a fixed number of
outputs per input and so requires a different design to ensure precise
signaling.  Our work therefore extends precise signaling capabilities
from regular to irregular streaming applications.  Like StreamIt, we
choose to send signals ``out of band,'' in our case via parallel
control edges, rather than attempt to enqueue them together with data
items.

Other streaming systems, such as Ptolemy~\cite{Ptolemy} and
Auto-Pipe~\cite{AutoPipe}, support a variety of dataflow semantics,
including multiple, differently-typed dataflow edges between a pair of
nodes. This support is in principle sufficient to implement a
control channel for signals.  However, except in restricted cases like
SDF, the systems do not specify how multiple channels between the same
two nodes are synchronized and so do not by themselves support precise
signaling.

Our work is influenced by a control messaging protocol developed by Li
et al.~\cite{PENG}.  That work, however, incurs additional complexity
to support asynchronous streaming dataflow and to impose well-defined
semantics on convergent dataflow edges in an DAG-structured irregular
application.  We preserve their idea of a credit protocol for
synchronizing data and control streams but realize this idea in a way
that is efficient for our target model and architecture.

The idea of processing part of a stream of items in a common context
is similar to facilities present in Apache Spark~\cite{Spark}.  Spark
streams consist of a sequence of RDDs~\cite{APACHE}, which are
discrete data sets, processed as a unit, that may contain multiple
elements.  Our abstraction supports some operations semantically
similar to Spark's but realizes them in the context of a single
wide-SIMD processor rather than the multicore and distributed systems
that Spark targets.

Other frameworks supporting streaming-like behavior, such as
CnC-CUDA~\cite{CNC}, utilize an ``in-band'' approach with
\emph{control collections} that mix control and data into a single
stream.  Control collections are analogous to our regions of items
with a common context.  Their implementation requires a tag for each
item to track the region associated with it.  In contrast, our
implementation keeps region boundaries synchronized with the data
stream without the need for tagging.  We compare these two approaches
in Section~\ref{secResults}.

\section{A Mechanism for Precise Signaling}

In this section, we describe how to synchronize data and control
signals between two successive nodes in a streaming pipeline.  The
mechanism bears some similarities to control flow in networking.  We
state the correctness properties of our design but relegate their
proofs to the appendix.

\subsection{Credit Protocol for Synchronizing Signals}

\begin{figure}[b]
	
\centering
\includegraphics[width=0.6 \columnwidth]{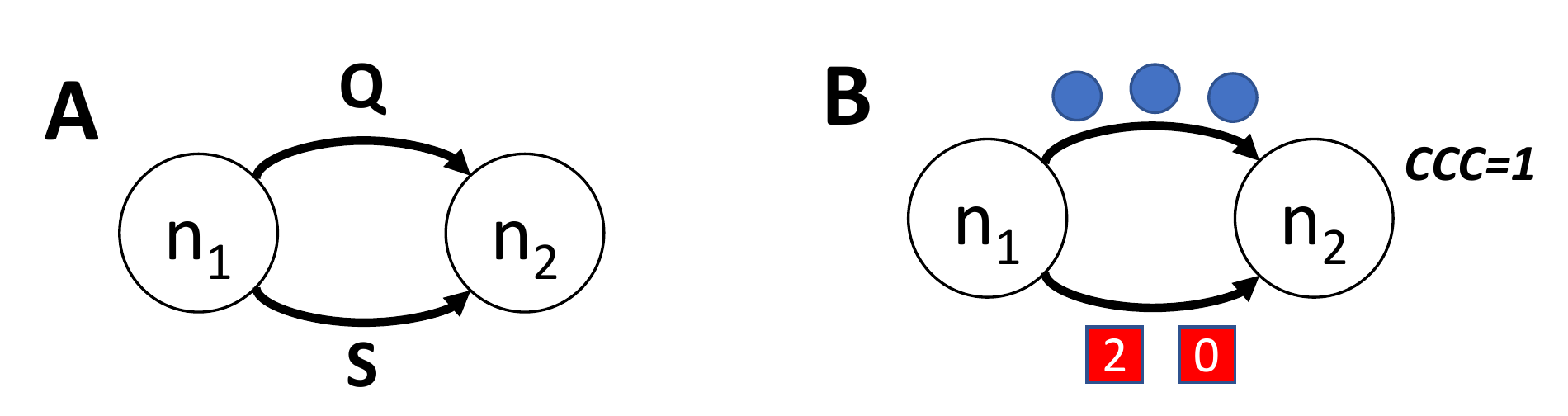}
	
\caption{(a) Two nodes with data and signal queues $Q$ and $S$ between
  them; (b) A possible state of the signal protocol, showing the
  credit associated with each signal (lower edge) and the current
  credit counter (right).  $n_2$ may consume one data item from $Q$
  before the first signal, then another two before the second signal.
\label{figSignals}}
	
\end{figure}

Let $n_1$ and $n_2$ be successive nodes in a pipeline, connected by
data queue $Q$. We add a separate, finite-sized \emph{signal queue}
$S$ between the nodes, as in Figure~\ref{figSignals}a.  Data
items are moved from $n_1$ to $n_2$ on $Q$, while signals are moved on
$S$.

We must ensure that, although data and signals move on separate
queues, their movement is synchronized so as to ensure precise signal
delivery. For this purpose, we introduce a \emph{credit protocol}
between $n_1$ and $n_2$. Each signal created by $n_1$ is assigned an
non-negative integer amount of \emph{credit}, which is transmitted
along with the signal on $S$.  Credit records a number of data items
that $n_2$ must process before it can receive the signal.

When $n_1$ emits a signal $s$, it uses two rules to set the credit
associated with $s$. (1) If no signal is currently queued on $S$, then
$s$ gets an amount of credit equal to the number of data items queued
on $Q$.  (2) If one or more signals are queued on $S$, let $s'$ be the
signal at the tail of $S$. Then $s$ gets an amount of credit equal to
the number of data items emitted by $n_1$ since $s'$ was enqueued.
Node $n_1$ maintains a counter of emitted data items, which is reset
each time it emits a signal, that is used to implement the second
rule.

The downstream node $n_2$ maintains a \emph{current credit counter},
initially set to 0, that tracks the number of items that can safely be
consumed before processing the next signal. Node $n_2$ uses the
following two rules to determine whether to process data or a signal
when it fires.  (1) If no signal is queued on $S$, $n_2$ may freely
consume any available data items on $Q$ without regard to the counter.
Otherwise (i.e., a signal \emph{is} queued on $S$), (2a) if the
current credit counter is non-zero, $n_2$ may consume only a number of
data items less than or equal to the value of this counter, which is
decremented once for each data item consumed.  (2b) If instead the
current credit counter is 0, let $s$ be the signal at the head of
queue $S$.  If $s$ carries more than 0 credit, that credit is removed
from $s$ and added to the current credit counter.  Otherwise, $n_2$
consumes $s$.

Figure~\ref{figSignals}b illustrates how the credit carried in the signals
and in the receiving node's credit counter synchronizes the two queues.

\begin{lemma}
\label{lemmaPrecise}
A signal $s$ emitted by $n_1$ is received by $n_2$ precisely when
$n_2$ has consumed all data items emitted by $n_1$ prior to $s$.
\end{lemma}

\subsection{Scheduling Applications with Signals}

A firing of a node $n$ proceeds in two phases: a \emph{data phase} and
a \emph{signal phase}.  In the data phase, $n$ consumes as many queued
data items as it can.  The number of items consumed is limited to the
minimum of three values: the number of queued items, the amount of
space in $n$'s downstream queue, and (if a signal is pending for $n$)
the value of $n$'s current credit counter.  Once $n$ can consume no
more data, if its current credit counter is 0, it enters the signal
phase, in which it consumes as many queued signals as it can.  Signal
processing ends when no queued signals remain, or when $n$'s current
credit counter becomes $> 0$ (and hence data must be consumed prior to
the next queued signal).

A node is \emph{fireable} if it has either data or a signal pending,
and if there is sufficient space in its output queue to hold any
outputs from the firing. The maximum number of output data items per
input item is known \emph{a priori} for each node, so the scheduler
can determine whether at least one data item can be consumed given the
available space on $n$'s downstream data queue. The maximum number of
\emph{signals} emitted per data item or signal input is also known
\emph{a priori}, so a similar determination can be made given $n$'s
downstream signal queue.  The scheduler repeatedly chooses some
fireable node and fires it until no node has queued data or signals
remaining.

\begin{lemma}
\label{lemmaDeadlock}	
Under the given firing/scheduling policy, an application pipeline
always finishes execution in finite time and so cannot deadlock.
\end{lemma}

\subsection{SIMD Extensions}

The above description assumes that nodes process input data items one
at a time.  However, on a SIMD-parallel processor, a node may process
an ensemble of multiple items at once.  Because a signal updates the
state of the receiving node, we must ensure that \emph{items appearing
  before and after a signal in the stream are not processed in the
  same input ensemble}.  Hence, if a signal is queued for a node, the
system must limit the size of the node's input ensemble to the value
of its current credit counter.  This requirement may adversely impact
SIMD occupancy if signals occur frequently; we study its impact in
Section~\ref{secResults}.

\section{Regional Context via Enumeration and Aggregation}

We now describe a developer-facing abstraction, \textit{enumeration
  and aggregation}, that allows application developers to describe
streaming computations in which regions of a stream must be processed
in a common context.  We have implemented this abstraction as an
extension to our MERCATOR system, building on the work of the previous
section.

Our abstraction assumes that regions of a stream with a common context
are represented as \emph{composite objects}, similar to the RDDs of
Apache Spark~\cite{Spark}.  The actual input provided to the
application is a stream of such objects.  Each object may contain zero
or more \emph{elements} of a common data type.  For example, an object
could be a line of text whose elements are characters, or a vertex
whose elements are its adjacent edges, or a list whose elements are
numbers.  Objects may contain different numbers of elements.

At a given point in the application's pipeline, the developer may
choose to ``open'' the stream of composite objects to create a stream
of all their elements.  We call this opening process
\emph{enumeration} of the objects.  The enumerated element stream
becomes the input to the next node in the pipeline.  In this and
subsequent nodes, the developer may access the \emph{parent object}
that gave rise to an input item to obtain context needed for
its processing.

The opposite of enumeration is \emph{aggregation}, which ``closes''
the context associated with a parent object.  The developer may choose
to emit a stream of results derived from individual elements, stripped
of their parent context, or to aggregate values computed from the
elements of each parent object (e.g.\ by summing them) and emit a
single result per parent.  Either way, the stream of results continues
down the pipeline.

\subsection{Developer Interface}

\begin{wrapfigure}[16]{L}{0.4 \columnwidth}
\includegraphics[width=0.9 \linewidth]{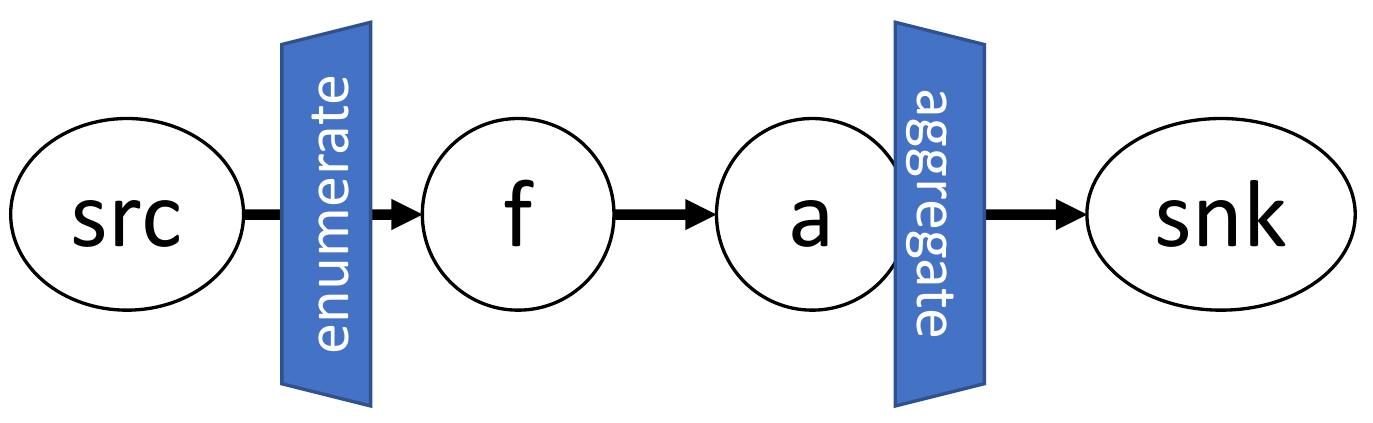}
    
\caption{A pipeline with enumeration and aggregation. The computation
  enumerates composite objects drawn from an input source, acts on
  their elements with a filtering node $f$, aggregates the filtered
  values in an accumulator node $a$, and writes the accumulated value
  from each object to an output sink.
      \label{figEnumApp}}
\end{wrapfigure}

To make these ideas concrete, consider the simple application whose
topology is sketched in Figure~\ref{figEnumApp}.  A stream of objects
of type \texttt{Blob}, each containing a collection of numbers, flows
from the source node.  The Blobs' elements are enumerated, and node
$f$ does some computation on each number in the element stream,
producing a (possibly shorter) output stream of numbers.  These
results are passed to node $a$, which sums the results from each Blob
and sends a stream of per-Blob sums to a sink node.

The listing of Figure~\ref{figSpecFile} specifies the application's
topology.  For each node, we specify the types of its input and output
streams.  The \texttt{enumerate} keyword at the input to node $f$
indicates that Blobs are to be enumerated starting there; subsequent
data types in the enumeration region of the pipeline are labeled
\texttt{from Blob}, indicating that items are to be processed in the
context of their parent Blobs.  Aggregation occurs at the output of
node $a$, where the \texttt{aggregate} keyword indicates that $a$
produces (up to) one \texttt{double} value per parent object rather
than per element.

Figure~\ref{figStubs} shows code to implement the application.
For each node, there is a \texttt{run()} function that processes items
in the node's input stream.  Output from a node is generated via the
\texttt{push()} function; because the application is irregular, not
every input might produce an output.  

The listing shows several functions specific to enumeration and
aggregation. \texttt{findCount()} is called once per
parent object to determine how many elements it contains.  The
\texttt{begin()} and \texttt{end()} functions, which may be defined
for each node receiving enumerated inputs, are executed before and
after the region of the stream associated with each parent object,
respectively.  The parent object associated with a node's current
input is accessible via \texttt{getParent()}.

Enumeration produces a stream of sequential \emph{indices} of elements
in each parent object. However, the application developer is
responsible for providing code to extract the elements from an object.
This design allows MERCATOR to remain ignorant of how objects are
organized internally.

\subsection{Implementation}

The MERCATOR system takes in an application topology, as shown in
Figure~\ref{figSpecFile}, and produces stubs for all the functions
shown in Figure~\ref{figStubs}.  The user then fills in the function
bodies with the actual code of the application.

We note that \emph{the code shown is CUDA, not C++}; hence, the
\texttt{run()} functions are actually called not with a single input
but with a SIMD ensemble of items in multiple threads, which execute
the function body in parallel for each item.  (The accumulation in
node $a$ would in practice be implemented atomically or with a
SIMD-parallel reduction.)  MERCATOR provides the runtime infrastructure
needed to transfer data from one node to the next and to schedule
nodes.

The signaling mechanism of the previous section is key to enabling
enumeration and aggregation.  At the point of enumeration, the runtime
generates a data stream of element indices together with signals
indicating the start and end of each parent object's elements.
Downstream nodes intercept these signals in order to update their
current parent object and to call the \texttt{begin()} and
\texttt{end()} stubs at the right times.  Because data before and
after a signal is never processed in the same SIMD ensemble,
operations on different parent objects' elements always happen in
separate calls to a node's \texttt{run()} function, and the result of
\texttt{getParent()} is the same for all items in an ensemble.

\begin{figure}[tb]
\centering
\begin{minipage}{0.4 \textwidth}
  \lstset{basicstyle=\scriptsize,breaklines=true}
  \begin{lstlisting}
Node src : Source<Blob>;
Node f   : enumerate Blob ->
  float from Blob;
Node a   : float from Blob ->
  aggregate double;
Node snk : Sink<double>;
      
Edges src -> f -> a -> snk;
  \end{lstlisting}
  \caption{Application topology specification illustrating enumeration
           and aggregation. \label{figSpecFile}}
\end{minipage}%
\hspace{0.05\textwidth}%
\begin{minipage}{0.5 \textwidth}
  \lstset{basicstyle=\scriptsize,breaklines=true}
  \begin{lstlisting}
void enumForF::findCount(Blob *b) 
{ return b->nElements(); }

void f::run(int i)
{
  Blob* b = getParent();
  float v = b->getItem(i);
  if (isGood(v)) push(3.14 * v);
}
	
void a::begin(Blob *b) { acc=0.0; }
void a::run(float v) { acc+=v; }
void a::end(Blob *b) { push(acc); }
  \end{lstlisting}
    \caption{Application code, with stubs generated from topology and
             developer-supplied function bodies. \label{figStubs}}
\end{minipage}

\end{figure}

\section{Results}
\label{secResults}

We implemented both our precise signaling infrastructure and the
enumeration and aggregation abstraction as extensions to our MERCATOR
framework and studied their performance on several benchmark
computations.  All experiments were conducted on an NVIDIA GTX 1080Ti
GPU (28 processors), using as many active blocks as could fit on the
device and a SIMD width of 128 threads per block.  Code was compiled
using CUDA v10 under Linux.  The abstraction penalty of the new
features was verified to be negligible in MERCATOR applications that
do not use them.

\paragraph*{Cost of Regional Context Abstraction} 
To characterize the performance impact of regional context, we began
with two simple benchmark computations.  Each benchmark operates on a
large array of integers in GPU memory, and each divides this array
into a series of regions.  The computation enumerates each region,
sums its elements, and produces a stream of per-region sums.  In the
first benchmark, the regions are of uniform size; in the second, the
size of each region is chosen uniformly at random between 0 and a
specified maximum.

Figures~\ref{figFixed} and~\ref{figVariable} show the time to process
an array of 512 million integers in each benchmark as a function of
the region size (fixed in the first figure, maximum in the second).
Focusing first on the test with fixed-sized regions, we see that
execution time decreases sharply as the region size grows from 32 to
the SIMD width of 128, then decreases more gradually for larger sizes.
This decrease reflects the lower frequency of region boundary signals
relative to the data stream as the region size increases.  For region
sizes on the order of several hundred of elements or more, the
abstraction overhead is small relative to the total cost of execution.

A second phenomenon observable in Figure~\ref{figFixed} is that the
overhead incurred by region boundaries changes non-monotonically with
region size.  In particular, overhead is locally minimized for region
sizes equal to a multiple of the SIMD width and then jumps sharply for
slightly larger sizes.  This behavior reflects the impact of region
boundaries on SIMD occupancy.  Recall that signals prevent elements in
two regions from being combined in the same SIMD ensemble, which is
required to ensure that each element contributes only to its own
region's sum. Region sizes that do not evenly divide the SIMD width
therefore require that nodes run with non-full input ensembles at
least once per region.  This loss of SIMD occupancy appears as reduced
application throughput.  For region sizes less than the SIMD width,
\emph{every} ensemble becomes non-full, which explains the large
performance impacts seen at region sizes below 128.

Figure~\ref{figVariable} shows a much-reduced impact of small size
variations on throughput.  Unlike the previous benchmark, but more
typically of real-world irregular applications, the region size is not
fixed. The sharp peaks of reduced throughput for worst-case region
sizes are therefore smoothed out, but the dominant effect remains:
larger region sizes incur less abstraction overhead.

\begin{figure}[tb]
\centering
\begin{minipage}{0.48\textwidth}
\resizebox{2.2in}{!}{
  \begin{tikzpicture}
    \begin{axis}[
	axis lines=middle,
	xmin=0,
	ymin=0,
	x label style={at={(axis description cs:0.5,-0.1)},anchor=north},
	y label style={at={(axis description cs:-0.1,.5)},rotate=90,anchor=south},
	xlabel=Region Size,
	ylabel=Execution Time (s),
	xticklabel style = {rotate=30,anchor=east},
	enlargelimits = false,
        xticklabel style={/pgf/number format/1000 sep=},
        xtick distance=256]
	\addplot[blue,thick,mark=square*] table [y=AvgTime,x=EnumSize]{syntheticFixed.dat};
    \end{axis}
  \end{tikzpicture}
}
  \caption{Execution time vs.\ region size \newline
           for sum app with fixed-size regions. \label{figFixed}}

\end{minipage}%
\vspace{0.08\textwidth}%
\begin{minipage}{0.48 \textwidth}
  \centering
  \resizebox{2.2in}{!}{
    \begin{tikzpicture}
	\begin{axis}[
	    axis lines=middle,
	    xmin=0,
	    ymin=0,
	    x label style={at={(axis description cs:0.5,-0.1)},anchor=north},
	    y label style={at={(axis description cs:-0.1,.5)},rotate=90,anchor=south},
	    xlabel=Max Region Size,
	    ylabel=Execution Time (s),
	    xticklabel style = {rotate=30,anchor=east},
	    enlargelimits = false,
            xticklabel style={/pgf/number format/1000 sep=},
            xtick distance=256]
	  \addplot[blue,thick,mark=square*] table [y=AvgTime,x=EnumSize]{syntheticRand.dat};
	\end{axis}
    \end{tikzpicture}
  }
  \caption{Execution time vs.\ max region size for sum app with variable regions. \label{figVariable}}

\end{minipage}
\vspace{-0.5in}
\end{figure}

\paragraph*{Comparison of Mechanisms for Communicating Context}
For our second experiment, we implemented a real-world application
taken from the DIBS benchmark set~\cite{DIBS}, a suite of applications
representative of data integration workloads.  The application, which
DIBS calls tstcsv-$>$csv but we refer to hereafter as ``taxi,''
processes a sequence of lines of text, each of which contains a tag, a
variable-length list of GPS locations specified as real-valued
coordinate pairs, and other data.  The goal is to parse each
coordinate pair, swap the elements of the pair, and emit the pair
together with the tag corresponding to its source line.

Our initial implementation of the taxi application operates on the raw
text in GPU memory.  It takes as input a stream of line start indices
and line lengths.  For each line, the first stage of the application
enumerates the line's individual characters as a stream, checks them
in parallel, and retains only those character positions (identified by
an open-brace character) that likely mark the start of a coordinate
pair.  The second stage verifies, again in parallel, that each
open-brace indeed marks a coordinate pair and, if so, parses the
pair's coordinates.  Each line's tag is parsed once when the line is
first enumerated and is then used to mark each parsed coordinate pair
for that line.

\begin{figure}

  \centering
  \resizebox{2.5in}{!}{
    \begin{tikzpicture}
      \begin{axis}[
	  legend pos=north west,
	  axis lines=middle,
	  xmin=0,
	  ymin=0,
	  x label style={at={(axis description cs:0.5,-0.1)},anchor=north},
	  y label style={at={(axis description cs:-0.1,.5)},rotate=90,anchor=south},
	  xlabel=Input size (Kilobytes),
	  ylabel=Execution Time (s),
	  xticklabel style = {rotate=30,anchor=east},
	  enlargelimits = false,
	  xticklabels from table={taxiRuntimeKB.dat}{NumCopies},xtick=data]
	\addplot[blue,thick,mark=square*] table [y=DoubleOpen,x=X]{taxiRuntimeKB.dat};
	\addlegendentry{Both stages enumerated}
	\addplot[red,thick,mark=triangle*] table [y=SingleOpen,x=X]{taxiRuntimeKB.dat};
	\addlegendentry{Stage 1 enumerated}
	\addplot[orange,thick,mark=x] table [y=Tagged,x=X]{taxiRuntimeKB.dat};
	\addlegendentry{Neither stage enumerated}
      \end{axis}
    \end{tikzpicture}
  }
  
  \caption{Execution time vs input size for three versions of the taxi app.
           \label{figTaxi}}
\end{figure}
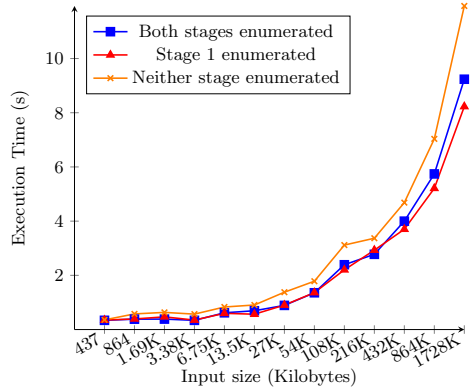

The first series of Figure~\ref{figTaxi} (square points) shows the
execution time of the taxi app as a function of its input
size. Larger file sizes were obtained by replicating the input file
from DIBS multiple times.  Mindful of the relationship between
abstraction penalty and region size, we then investigated how the
input data in the taxi app determined SIMD occupancy.  Input lines
have an average length of 1397 characters, so regions corresponding to
each line in the stage 1 are large, and the penalty to occupancy is
expected to be low.  In contrast, lines contain on average only 45
coordinate pairs, less than the SIMD width, and so would be expected
to incur a large penalty to occupancy in stage 2, whose region size
is determined by the number of pairs per line.  Indeed, we found
that stage 1 was fired with full SIMD ensembles 91\% of the
time, while stage 2 had full ensembles only 9\% of the time.

When regional context changes frequently relative to the SIMD width,
the occupancy cost to performance of our implementation may exceed the
cost (mainly extra memory accesses) of replicating this context along
with every data item.  We therefore developed a second version of the
taxi app that used enumeration to provide context in stage 1 but
explicitly marked each open-brace with its line's tag before sending
it to stage 2.  The latter stage does not utilize the enumeration
abstraction and so can process items from multiple lines in one
ensemble, achieving essentially full SIMD occupancy. The second series
in Figure~\ref{figTaxi} (triangular points) shows that improved
occupancy results in lower total execution time.  However, using the
same strategy to tag each character of each line in stage 1, while it
slightly improves occupancy by avoiding enumeration entirely, incurs
substantially more overhead due to the much greater number of elements
to be tagged per region.  The third series (x points) shows that, at
the largest input size tested, a pure tagging implementation is
roughly 30\% slower than one that judiciously uses either tagging or
our design as appropriate for each stage.

We conclude that the best way to provide regional context to streaming
applications on a SIMD architecture depends strongly on the
performance tradeoff between reduced SIMD occupancy and reduced
representation overhead.  Each stage of a pipeline may represent a
different point in this tradeoff, and the highest-performing
implementation (dense or sparse) for regional context may therefore
vary between stages.  Ultimately, this choice should be made
transparently to the application developer based on profile-guided
feedback.

\section{Conclusion and Future Work}

We have described an abstraction, enumeration and aggregation, to
support stateful streaming computation based on regional contexts.  We
presented an implementation of this abstraction for irregular
streaming computations on SIMD-parallel architectures such as
GPUs.  Our abstraction relies on a sparse implementation of precise
signal delivery between computational stages.

We characterized the cost of the abstraction on benchmark computations
and demonstrated that the best strategy for realizing it may depend on
the relationship between region size and the architecture's SIMD
width.  Future work will include more careful modeling and/or
empirical measurement of the costs of alternative implementations,
with an eye toward allowing the MERCATOR runtime to transparently
choose between strategies based on the typical number of elements per
region.

Another direction for future work will investigate how to lower the
abstraction penalty of precise signaling for SIMD occupancy. When the
effects of a signal on a node's state are limited and well-defined
(e.g.\ changing the parent object pointer), the node may be able to
compute the correct state (pre- or post-signal) to expose to the item
in each SIMD lane separately.  Computing the correct state per item in
each node, rather than storing it with items in the queues between
nodes, would offer the same efficient representation of state as in
our design while eliminating signals' cost to SIMD occupancy.

\subsection*{Acknowledgments}

This work was supported by NSF CISE awards CNS-1763503 and
CNS-1500173.

\bibliographystyle{splncs04}
\bibliography{Timcheck_Buhler_Signals}

\clearpage

\section{Appendix: Proofs Omitted in Text}

\subsection{Proof of Lemma~\ref{lemmaPrecise}}

\begin{proof}
	We proceed by induction on the number of signals already on the signal
	queue $S$ when $s$ is emitted.
	
	Suppose $S$ is empty when $n_1$ emits $s$. All items emitted by $n_1$
	prior to $s$ either have been consumed by $n_2$ or are present on $Q$.
	The protocol assigns $s$ an amount of credit equal to the size
	of $Q$. This amount is then added to $n_2$'s current credit counter,
	which was previously 0.  Finally, $n_2$ consumes $s$ precisely when
	its current credit counter returns to 0, which happens once $n_2$
	consumes the items that were present on $Q$ when $s$ was emitted.
	
	Now suppose $s$ is emitted when $S$ is not empty.  $s$ receives a
	number of credits equal to the number of items added to $Q$ since the
	prior signal $s'$. We know inductively that $n_2$ consumes $s'$
	precisely when all data items emitted prior to $s'$ have been
	consumed.  At this point, the only items on $Q$ must be those emitted
	after $s'$ but before $s$, and $n_2$'s current credit counter is 0
	since it just consumed a signal.  Conclude that $n_2$ will transfer
	the credit in $s$ to its current credit counter and will then consume
	exactly those data items emitted after $s'$ but before $s$ before it
	consumes $s$ itself.
\end{proof}

\subsection{Proof of Lemma~\ref{lemmaDeadlock}}

\begin{proof}
Our proof of deadlock-freedom relies on the following two claims.

\begin{claim}
A node cannot have a current credit counter $> 0$ without a pending
data item.
\end{claim}

\textit{Proof of claim}:
Suppose that a node's current credit counter is $> 0$. The credit in
the current counter was transferred from the signal $s$ currently at
the head of the signal queue; it cannot remain from prior signals
because the node did not even check for $s$ until its current credit
counter last became 0. Hence, this credit was assigned to $s$ to cover
data items that were enqueued at the time that $s$ was issued.  Since
not all credit has yet been consumed, at least one of these items
is still enqueued.

\begin{claim}
If a node $n$ has either pending data or a pending signal, one of the
following holds: (1) $n$ can consume a data item; (2) $n$
can consume a signal; (3) $n$ is blocked due to insufficient
space in its downstream queues.
\end{claim}

\textit{Proof of claim}:
The node either has credit or not.  If it has credit, then by the
previous claim it has pending data that can be consumed.  If it has no
credit but has a pending signal, then either the signal can be
consumed, or credit can be transferred from the signal; in the latter
case, there must again be data corresponding to this credit.  If there
is no credit and no pending signal, then there must be pending data,
which can be consumed without credit in the absence of pending signals.
In all cases, the node can consume some input unless its downstream
queues lack sufficient space.

We now proceed to prove the original lemma.  If any node in the
pipeline has pending data or a signal, then let $n$ be the last such
node.  Either $n$'s downstream queues are empty (else its successor
would have pending data or signals), or $n$ has no successor, i.e., it
is the last node in the pipeline, which has unbounded output space and
cannot block. Hence, node $n$ is not blocked on its downstream queues
and so, by the second claim, can be fired to consume input.
	
We conclude that the application terminates only when all nodes have
exhausted their inputs.
\end{proof}

\end{document}